# Hot carrier dynamics in a dispersionless plasmonic system

Banoj Kumar Nayak[1], S. S. Prabhu[1], Venu Gopal Achanta[1*]

[1]Tata Institute of Fundamental Research, Homi Bhabha Road, Mumbai 400005 INDIA

Email : achanta@tifr.res.in

**Abstract:** Hot carrier dynamics in a broadband, dispersionless plasmonic structure are studied by pump-probe measurements with 45 fs time resolution. The role of direct excited as well as plasmon generated hot carriers on low energy probe plasmons are studied by simultaneous measurement of differential transmittance and reflectance. While the pump fluence dependence on the decay times is linear for hot electrons and plasmon generated hot electrons, when pump is near resonant with the X-symmetry point, decay time showed quadratic dependence on the pump fluence. Decay times of 800 nm degenerate pump-probe measurements highlight the difference in surface (reflection) and the bulk (transmission) mechanisms. Decay time corresponding to the hot carrier relaxation is in the 1 -3 ps range for different excitation energies. Rise time, governed by the plasmon to hot carrier conversion and electron - electron scattering processes, is about 200 ± 35 fs for the hot carrier and hot plasmon excitation cases which increased to about 485 ± 35 fs for when pump is resonant with interband transition at X- symmetry point. Though plasmonic system helps excite hot carriers that are not directly accessible in a bulk metal, results show that the dynamics are governed by the bulk metal band structure. The dipole matrix element for each of the transitions is estimated by density matrix calculations.

## Introduction:

Surface plasmon polaritons (SPPs) are coherent oscillations of free electrons resulting in delocalized charge density wave at the metal-dielectric interface [1]. Metal-dielectric structures that support SPPs are studied for many diverse applications like nanolasers, superlens, second harmonic generation, extraordinary transmission among others [2]. Plasmon propagation length, lifetime, effect of plasmons on a short laser pulse and modulation of plasmons are widely studied [3-5]. Plasmon dynamics are governed by plasmon decay into high energetic (hot) carriers that can result in electrons and holes or heat depending on the energy [6-8]. The role of hot electrons in diverse phenomena that include plasmon - exciton coupling, modulation of Raman and other nonlinear optical phenomena, hot carrier injection, local doping, local heating, chemistry among others are of contemporary interest [9-13]. Hot carrier injection and their dynamics in metal nanoparticle - semiconductor complexes, especially, received considerable recent attention [8, 13,14]. The energy, lifetime and population of these hot carriers play major role in transport mechanism and thus the plasmon generated hot electron dynamics are important. As the carrier distribution changes by generation of hot electrons in sp- band and d- band holes, optical properties are modified at ultrafast time scales. Plasmon decay to hot carriers can occur through direct interband transitions and/or intraband transition assisted by phonons and geometry of the pattern [6, 9, 15]. Direct interband transition takes place when photon energy is higher than the energy difference between d-band and Fermi level in gold. An example of this is, 5d – 6sp transition at L-symmetry point in Gold. When the plasmon energy is less than the interband threshold energy, intraband transitions take place, where 6sp electrons are excited above the Fermi level. It is also possible that the SPPs through the geometry of the plasmonic structure and phonons provide the required additional momentum for indirect transitions in the parabolic sp-band [6, 14, 16, 17]. Thus, due to strong localized field that has large gradients, conventional dipole forbidden processes may take place in plasmonic structures. In addition, the flatter d-band means large electron density of states (DoS) near the symmetry points like at X- and L- points. This large DoS allow d- to sp- transitions at the symmetry points though the transition matrix elements are weak and thus the metal band structure contributes to the hot carrier dynamics. Theoretically, it has been shown by Sundararaman et al that the hot carrier generation due to plasmon decay is governed by the underlying bandstructure [7].

While the initial electron thermalization due to electron – electron interaction takes place on the timescales of the order of 100 fs, the excess energy, compared to the lattice, of the hot electrons is dissipated as electron – phonon interaction over picoseconds [18-23]. This process of fast thermalization of electrons followed by cooling via lattice interaction is explained by a two temperature model (TTM) [24]. TTM explains the room temperature dynamics quite well though discrepancy between the low temperature measurements and TTM was reported in gold films due to non-thermal electron distribution when metal is heated by a laser pulse [25]. Carrier dynamics in various bulk metals have been studied by different techniques like photoelectron spectroscopy, photothermal modulation, and pump induced reflectance and transmittivity [26-29]. An improvised TTM was proposed recently to account for the creation of non-thermal electron distribution at sub-100fs time scales when excited by an ultrashort laser pulse [30].

Metal nanoparticles and plasmonic crystals offer discrete resonances specific to the structures being studied [5, 18, 31]. The dynamics are limited by the geometry defined plasmon dispersion [4]. A decay time of 0.6ps for hot electrons is reported in one-dimensional plasmonic grating structure [5]. Other plasmonic device structures have also reported a lifetime less than 1ps [13]. Previous studies on metal nanoparticles reveal that the relaxation time of these hot electrons strongly depends on the probe wavelength and this process is observed to be slower at the localized plasmon resonances [32, 33]. It has also been observed in a plasmonic crystal (1-d grating) that the differential transmission changes by upto 3% when pump and probe are resonant with SPP mode [5]. Though the patterning of metal has been shown to effect the absorption thus making it independent of the bulk metal, carrier dynamics are not studied in a system which has dispersionless plasmonic response. To show the role of metal band structure on the plasmon mediated hot carrier generation, we may consider, either a dispersionless plasmonic system or a structure in which geometry independent plasmon dispersion is possible.

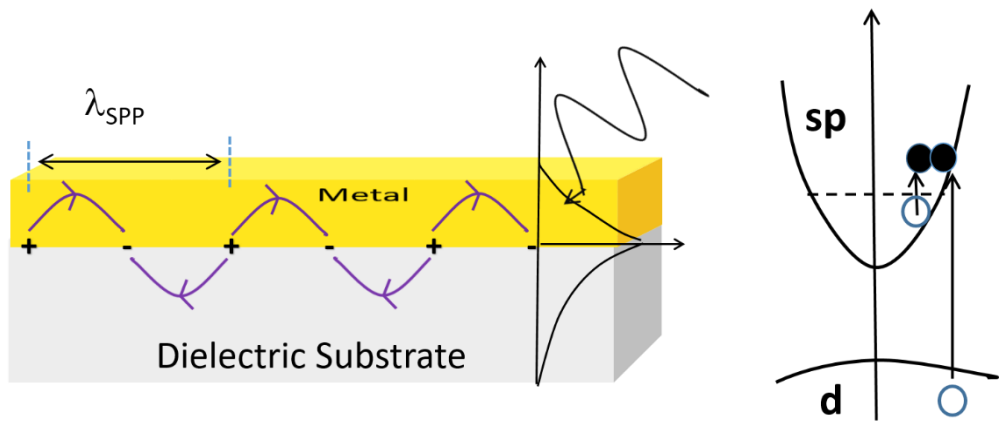

Fig. 1 Schematic of a metal-dielectric structure showing the optical excitation of surface plasmon polaritons (SPP) at the interface. Evanescent decay of surface mode is indicated by the decay profiles into the metal and dielectric layers. SPP mode excites interband (long arrow from d- to sp- band) or intraband (short arrow indicating transition from sp-band to above Fermi energy) transitions. Such hot carriers decay by the two temperature model involving electron – electron and electron – phonon scattering.

Plasmonic quasicrystals (PlQC) are quasiperiodic array of metal – dielectric structures either in the form of sub-wavelength metal scattering elements or air holes in metal film [34]. They offer several advantages that include broadband SPPs that are independent of the launch angle and polarization. The large local field enhancement results in nonlinear optical response that include phase matching independent generation of second harmonic from thin metal films [35]. Their unique features and possible applications are discussed in a review article [36]. It would be interesting to study the hot carrier dynamics in dispersionless plasmonic structures.

In this paper, we present simultaneous differential reflection and transmission measurement results on a broadband, dispersionless PlQC structure [34, 35]. In the degenerate and non-degenerate pump – probe configurations, we have studied the effect of hot electrons generated by free electrons, non-resonant and resonant plasmons on probe generated low energy plasmons. While the excited free carriers decay according to the TTM as shown in the schematic in Fig. 1, excited plasmons excite interband or intraband transitions generating hot carriers which also follow the TTM. Resonant plasmons excited by the pump and probe beams elucidate the plasmon-plasmon interaction. This paper is organized as follows. We first describe the experimental details including the choice of different pump and probe wavelengths followed by the plasmonic quasicrystal sample fabrication and plasmonic response. We will then present the pump-probe experimental results and discussion. The results are summarized in Conclusions.

## Experimental Details:

As the pump excited carrier dynamics modulate the electronic energy distribution, transmission and reflection response of plasmonic structure change as a function of time. These changes are monitored by their effect on a probe pulse that excites low energy plasmons. For these measurements, a 2-color pump-probe experiment was setup for simultaneous differential reflection and transmission measurements. A 3 mJ amplifier laser (Spectra Physics, Spitfire) at 800 nm with 25 fs pulse duration at 1 kHz repetition rate was used to pump TOPAS Optical Parametric Amplifier. A combination of 2 TOPASs or one TOPAS output and 800nm-amplifier output are used as Pump and Probe beams, respectively in the two-color pump probe setup. The zero delay between the pump and probe beams is identified by measuring the transient reflectivity change in GaAs. Path length variation between pump and probe beams is achieved by standard delay line. The collinear pump and probe beams are focused with a 10 cm lens and spatially overlapped at the focal plane where sample was mounted. Differential transmission and reflection are measured simultaneously using a combination of photodiodes (Hamamatsu) and lock-ins (SRS830) that are locked to the chopping frequency of the pump beam. A suitable combination of interference filters (Semrock) were used to filter out pump beam from being detected. Differential transmission and reflection signals are given by $\Delta T = (T_P - T_0)$ and $\Delta R = (R_P - R_0)$, respectively, where $T_P$ is the probe transmission with Pump beam ON and $T_0$ is transmission of probe without the pump beam. Similarly, $R_P$ and $R_0$ are the reflected probe intensities with and without the pump beam, respectively. Normalized with the $T_0$ (or $R_0$) and measured as a function of delay between the pump and probe pulses gives the differential transmittance (reflectance).

In our studies, the probe wavelength was set to be 800nm to excite low energy plasmons in the PlQC structure. We have chosen different regimes of pump wavelength depending upon carrier generation mechanism. In the first case, we generate free carriers by pumping with 500 nm (2.48 eV) above the threshold energy of interband transition at L-symmetry point ($\Delta_L$). In the second case, we use high energy plasmon excitation below the threshold energy of interband transition such that intraband transition from sp-band to above Fermi level takes place with 575 nm (2.156 eV) pump. In third case, we use plasmons close to the interband transition at X-symmetry point ($\Delta_X$) to study resonant effects, if any, with pump at 650 nm (1.908 eV). In the fourth case, degenerate pump – probe measurements with low energy pump plasmons resonant with the probe plasmons are carried out with 800 nm pump (1.55 eV).

The plasmonic structure studied is a plasmonic quasicrystal made of quasiperiodic array of air holes conforming to 5-fold rotation symmetry in a metal film over a substrate. Figure 2(a) shows the scanning electron micrograph (SEM) image of quasiperiodic air hole arrays fabricated by e-beam lithography on 50 nm gold film sputtered on a glass substrate. It has a broadband plasmonic response spanning from 550nm to 1150 nm wavelength as shown by the enhanced transmission spectrum (ratio of transmission through patterned and unpatterned gold) of Fig. 2(b). Dispersionless nature of the plasmons in such structure was reported earlier [34, 35]. The short range separation between nearest neighbor air holes in the quasi periodic array is about 10 nm and the long range periodicity is 600 nm. The dense k-space supports polarization independent and dispersionless plasmon band with phase-matching independent nonlinear optical properties [34, 35].

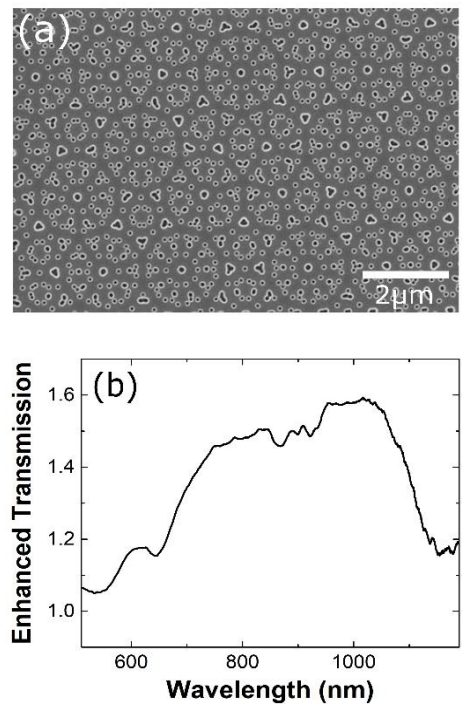

Fig. 2 Scanning electron microscope image of the quasiperiodic pattern of air holes on a gold film is shown in (a). (b) shows the measured transmission enhancement spectrum with respect to unpatterned gold film. Broadband plasmon mediated enhanced transmission is seen in the 550 nm to 1150 nm wavelength region.

## Results and Discussion:

Differential transmittance and reflectance of 800 nm (1.55 eV) probe beam as a function of time delay with respect to the pump pulse centered at different wavelengths mentioned above was measured. In a two temperature model, the rise ($\tau_r$) and decay ($\tau_d$) times relate to the electron – electron and electron – phonon interactions, respectively [27,31,37,38]. With the contribution from change in lattice temperature due to residual carriers considered to be a constant (A), the temporal profile is given by a convolution of Gaussian pulse and the function,

$$\Delta T = A + [B.e^{-\frac{t}{\tau_d}}.(1 - e^{-\frac{t}{\tau_r}})] \tag{1}$$

where A and B are constants. The rise and decay time constants for different pump wavelengths and excitation fluences are extracted by deconvoluting with the Gaussian laser pulse profile.

Density matrix calculations were carried out for each of the 4 pump wavelength cases. It should be recalled that different pump wavelengths correspond to L-symmetry point 5d-6sp interband excitation (500nm), plasmon excitation resulting in 6sp intraband excitation at L-symmetry point (575nm), X-symmetry point 5d-6sp interband and L-symmetry point 6sp intraband excitation which are both plasmon mediated (650nm) and L-symmetry point plasmon mediated intraband excitation (800nm). As change in reflectivity is related to the change in refractive index which is related to the carrier density, for each transition we calculated the modulation in carrier density (ΔN/N) and compared it with the measured ΔR/R. Further, in the density matrix model, by taking the carrier lifetimes to be those that are obtained by fitting the experimental data with the 2-temperature model, the dipole matrix element for different transitions are estimated. In the density matrix model, the diagonal terms for the population are given by the following expressions,

$$\frac{d\rho_{11}}{dt} = \frac{1}{i\hbar}\sum_{k} H_{1k}\rho_{1k} - \rho_{1k}H_{k1} + \frac{\rho_{33}}{\tau_3} \qquad (2)$$

$$\frac{d\rho_{22}}{dt} = \frac{1}{i\hbar}\sum_{k} H_{2k}\rho_{k2} - \rho_{2k}H_{k2} - \frac{\rho_{22}}{\tau_2} \qquad (3)$$

$$\frac{d\rho_{33}}{dt} = \frac{1}{i\hbar}\sum_{k} H_{3k}\rho_{k3} - \rho_{3k}H_{k3} - \frac{\rho_{33}}{\tau_3} + \frac{\rho_{22}}{\tau_2} \qquad (4)$$

where, $\tau_2$ and $\tau_3$ are the relaxation times of levels 2 and 3, respectively. $H_{ij}$ is the interaction Hamiltonian given by the product of dipole matrix element ($\mu_{ij}$ for the *ij* transition) and the electric field (E(r,t)) that depends on the incident pulse. The non-diagonal elements of density matrix for $\tau_{mn}$ being the dephasing time constant are given by,

$$\frac{d\rho_{mn}}{dt} = \frac{1}{i\hbar}\sum_{k} H_{mk}\rho_{kn} - \rho_{mk}H_{kn} - i\frac{E_{mn}}{\hbar}\rho_{mn} - \frac{\rho_{mn}}{\tau_{mn}} \qquad (5)$$

To briefly outline the results, the hot carrier excitation and hot plasmon excitation show similar fast decay times showing that the plasmon to hot carrier conversion is very fast and carrier - phonon scattering mechanism is the dominant decay mechanism as expected. It would be interesting to note that the decay time as a function of excitation fluence vary linearly for 500 nm, 575 nm and 800 nm pump but has a quadratic dependence for 650 nm excitation wavelength (Figs. 3c, 4c, 5c, and 6c). In the following, we present detailed results for each of the excitation wavelengths followed by discussion of the results.

From the data taken for higher fluences, there is a residual part in differential reflectance even after non-equilibrium electron decay through electron-phonon coupling. This could be attributed to lattice contribution as the lattice temperature rises [27,39,40]. The change in lattice temperature can be calculated from the measured ΔR/R at different excitation wavelengths and thermo-reflectance coefficient ($C_{TR}$) at 800 nm of 3.2e-5 [41-42]. The ΔR/R values are 9.6x10$^{-4}$, 1.27 x10$^{-3}$, 9.73 x10$^{-4}$, and 4.84 x10$^{-4}$ for excitation wavelength of 500nm at fluence 15mJ/cm$^2$, 575nm at 46mJ/cm$^2$, 650nm at

14mJ/cm$^2$, and 800nm at 10mJ/cm$^2$, respectively. From these values, the calculated lattice temperature change is 30 K, 40 K, 30 K and 15 K for excitation at 500nm, 575nm, 650nm and 800nm, respectively.

A comparison of the differential transmittance data at different excitation energies show that, the maximum modulation of about 1.7% in transmittance is achieved when free carriers are excited by the pump. Unlike Ref. 5, where the maximum change is observed in reflectance when the probe is at SPP resonance, present data shows weaker modulation of reflectance compared to transmittance. In the following the pump fluence dependent dynamics for each excitation wavelength are presented.

**Results of 500nm pump:**

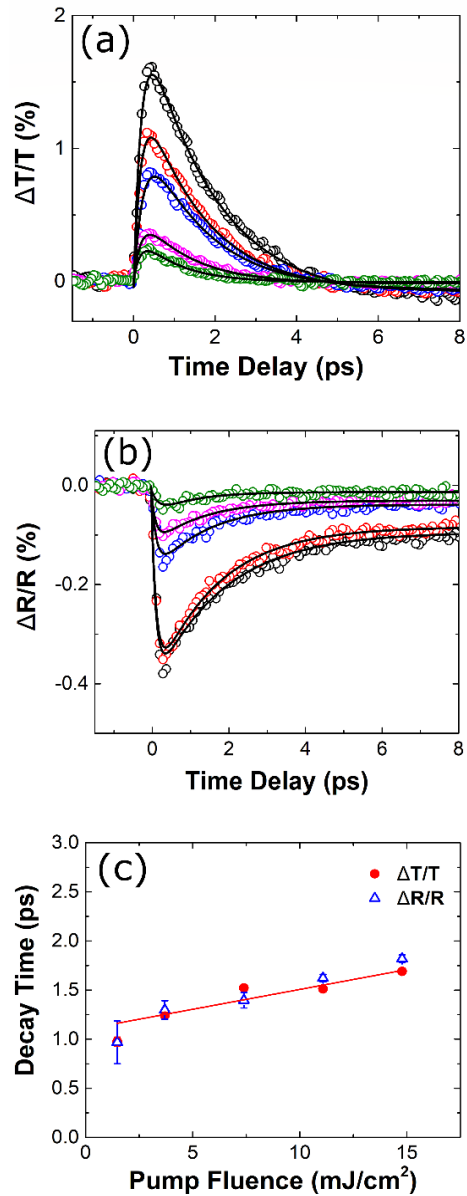

Fig.3 Differential transmission (a) and reflection (b) decay scans at different pump fluences are shown along with fits. Pump wavelength is 500 nm and probe is at 800 nm. (c) shows the pump fluence influence on the decay time of differential transmissivity (Red circles) and reflectivity (Blue triangles) data. Solid line shows the linear fit.

Figure 3 shows the percentage change i.e. ΔT (Fig.3a), ΔR (Fig.3b) decay profiles and the decay time dependence on the pump fluence (Fig.3c) for the pump and probe wavelengths set to 500 nm and 800 nm, respectively. The rise and decay times of each of the decay plots are tabulated in Table 1. The optical response of the material changes as the electron energy distribution and the optically excited carrier density evolves. Since we are exciting at 500 nm (2.48 eV) which is above the threshold of interband transition, direct transition occurs from d-band to above Fermi level generating high energy carriers. They thermalize through electron-electron scattering, which is proportional to $(E-E_F)^2$, to form a hot carrier distribution resulting in a rise in electron temperature [6, 21, 28, 31, 32]. Thermalization time is, typically, of the order of few hundreds of femtoseconds. This process is responsible for rise of the pump-probe signal [19, 39].

While the pump fluence dependence on the rise time (Table 1) showed that transmission rise time is within the 200 ± 8 fs for all pump fluences, the rise time of the reflection data was found to be in the range of 136 ± 12 fs. This difference in the transmission and reflection rise times could be explained based on the fact that, the carriers near the surface (within the skin depth) affect the reflection but the transmission is affected by the propagation inside the metal film. Skin depth at 500 nm (pump) is about 45 nm but that at 800 nm (probe) is about 25 nm.

After thermalization, the hot electrons cool by transferring their energy to the lattice by electron-phonon interaction. The phonon energy is, typically, of the order of several meV [43] and thus the energy loss of a hot electron in an electron-phonon collision is low compared to its excitation energy [9]. So, the decay process is slower compared to rise time in pump-probe signal. The relaxation process happens over a timescale of the order of a few picoseconds. Figure 3c shows that the decay times governed by the electron – phonon scattering are similar for both reflection (Fig. 3b) and transmission (Fig. 3a) data as these are governed by the lattice interactions. Further, it may be noted that with the pump fluence, the decay times increase linearly as shown in Fig. 3c. From the decay times, using the density matrix model, we estimated the dipole matrix element to be 1.32 ±0.02 x $10^{-31}$ C.m.

Table 1: Excitation wavelength- 500nm, Probe wavelength-800nm. Errors shown are the numerical fit errors which are much smaller than the pulse width of 35 fsec.

| Excitation power (mJ/cm$^2$) | ΔT Rise time (ps) | ΔT decay time (ps) | ΔR Rise time (ps) | ΔR decay time (ps) |
|---|---|---|---|---|
| 15 | 0.207±0.01 | 1.690±0.02 | 0.124±0.01 | 1.820 ± 0.04 |
| 11 | 0.199±0.01 | 1.512±0.02 | 0.137±0.01 | 1.620 ± 0.04 |
| 7 | 0.276±0.01 | 1.522±0.02 | 0.145±0.02 | 1.396 ± 0.08 |
| 4 | 0.206±0.01 | 1.240±0.03 | 0.130±0.02 | 1.23 ± 0.09 |
| 1 | 0.192±0.02 | 0.975±0.05 | 0.145±0.11 | 0.97 ± 0.22 |

**575nm pump:**

Pump and probe wavelength are set at 575nm and 800nm, respectively. Differential transmission (Fig. 5a) and differential reflection (Fig. 5b) are shown for different pump fluences along with fits based on 2-temperature model. Transmission and reflection decay times are identical and have a linear dependence on the pump fluence (Fig. 5c). The laser spot size was smaller at 575 nm and thus the fluence values are larger than that at other wavelengths. With decrease in pump power, the differential signal decreases as expected and the decay time is shorter significantly. For all the excitation pump fluences, the rise time of differential transmittance signal is in the range 199 ± 25 fs and the differential reflectance rise time is in the range 160 ± 22 fs. The rise times are thus within one pulse width. Table 2 summarizes the rise and decay times for different pump powers.

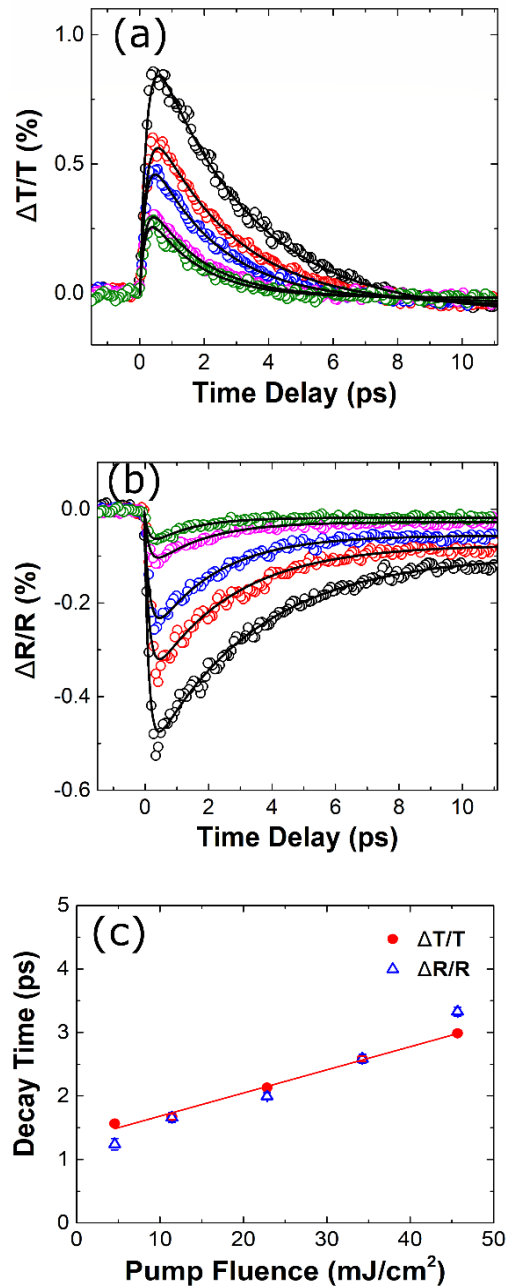

Fig.4 Differential transmission (a) and differential reflection (b) as a function of delay for different pump fluences. Pump wavelength is 575 nm and probe is at 800 nm. Decay time of transmission and reflection data as a function of pump fluence are shown in (c). Solid red line shows a linear fit.

At 2.16 eV (575nm) pump, the excitation energy is lower than the gold interband transition threshold at ~2.4eV near L-symmetry point [44-46]. Since the quasiperiodic structure has broadband plasmonic structure from 550nm to 1000nm, the 575nm pump excites the high energy plasmon modes. The surface plasmon decay leads to hot carriers which decay via the sp-band of gold with the initial and final energy states having different crystal momenta [16]. Difference in momentum is provided by phonons and geometry of the plasmonic structure [6, 14, 16-18]. Reflection decay times are slightly longer than the transmission decay times and both increase as a function of excitation pump fluence. This slower decay with increasing pump fluence could be due to screened coulomb potential at higher excited carrier densities. As mentioned earlier, the skin depth at 575nm (pump) is about 30 nm but that at 800 nm is about 25 nm. Thus, the transmitted carriers undergo scattering in the bulk like gold with electrons, phonons, defects as well as affected by the nanopattern. But the reflected probe beam plasmons at L-point are effected only by the surface carrier density. The dipole matrix element estimated for this process is 1.44 ±0.01 x $10^{-31}$ C.m.

Table 2: Excitation wavelength- 575nm, Probe wavelength-800nm

| Excitation power (mJ/cm$^2$) | ΔT Rise time (ps) | ΔT decay time (ps) | ΔR Rise time (ps) | ΔR decay time (ps) |
|---|---|---|---|---|
| 46 | 0.217±0.01 | 2.987±0.04 | 0.138±0.01 | 3.33±0.07 |
| 34 | 0.224±0.01 | 2.586±0.04 | 0.161±0.01 | 2.59±0.07 |
| 23 | 0.191±0.01 | 2.131±0.04 | 0.182±0.01 | 2.00±0.05 |
| 11 | 0.188±0.01 | 1.674±0.04 | 0.177±0.02 | 1.66±0.08 |
| 5 | 0.174±0.01 | 1.566±0.05 | 0.165±0.03 | 1.24±0.09 |

**650nm pump:**

In the 3$^{rd}$ set of measurements, pump and probe wavelengths were set to 650nm and 800nm, respectively. The decay time of differential transmissivity and differential reflectance for 14 mJ/cm$^2$ average pump fluence was found to be 1.82 ps and 1.73 ps, respectively. With decrease in pump power, the differential signal decreases and the decay time also got shorter but rise time of signal did not change much. The rise and decay times are summarized in Table 3 with the error bars obtained by minimizing the $\chi^2$ during the fits. From the range of values obtained for different pump fluences, the error in estimated life times are within one pulse width (45 fsec).

Rise time of the transmission data for all pump fluences is in the range of 486 ± 43 fs and that for the reflection data is in the range of 227 ± 59 fsec. These rise times are almost a factor of 2 larger compared to those at 500 nm and 575 nm excitation. The shorter reflection data rise time with respect to transmission rise time is consistent with the fact that the decay of free carriers at the surface influence the reflection data while the transmission is affected by the scattering in the metal film. 650 nm or 1.907 eV is between the interband transitions at the X- and L-symmetry points in gold [47, 48]. The dynamics are thus expected to be different from those at L-symmetry point that are excited at other pump energies.

It would be interesting to note that both the transmission and reflection decay times vary quadratically with the pump fluence unlike the linear dependence seen at 500 nm and 575 nm pump excitations. Such a behavior could be associated with multiple excitation pathways. In gold band structure, the presence of X- and L-symmetry points was reported by ellipsometric measurements as well as from photoluminescence measurements [47, 48]. Recently, SPP modified photoluminescence from patterned gold films were shown to be dominated by the X- symmetry point 5d – 6sp transition when the emission energy is in between $\Delta_X$ and $\Delta_L$ energies [44]. Though the transition matrix element for to 5d- to 6sp-band transition at X- symmetry point is much weaker than that at L-point, the increased DoS and plasmon mediated strong local field lead to hot carrier generation at X-point. Quadratic dependence of decay time on the fluence as well as plasmon mediated hot carrier injection at X-symmetry point could be useful for photovoltaic and other applications where carrier extraction as well as injection at specific energies is required [8, 13,14].

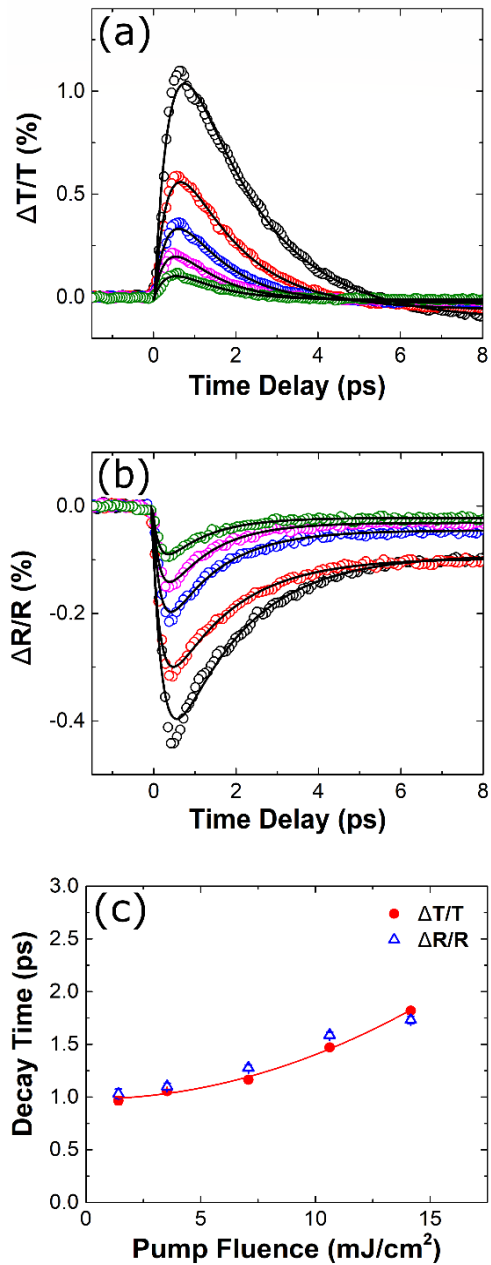

Fig.5 Differential transmission (a) and reflection (b) decay scans at different pump fluences are shown along with fits. Pump wavelength is 650 nm and probe is at 800 nm. (c) shows the pump fluence dependence of the decay time of transmission (Red circles) and reflection (Blue triangles) data. Solid line shows the quadratic fit.

In the present case, the quadratic dependence of the decay time corresponding to the probe beam (at 800 nm or 1.55 eV) generated low energy intraband carriers in the sp-band at the L-point on the pump fluence could be explained as follows. At low pump power, both X-valley transition and L-valley transition contribute to carrier generation. However, carrier relaxation at X-valley will not affect the dynamics of plasmon mediated low energy carriers created by the probe beam at the L-valley. Therefore, at low pump powers, decay lifetime will be less as fewer carriers are excited at L-symmetry point. However, at higher fluences, the number of carriers getting excited at the L-symmetry point get significant and comparable to the other cases and thus the decay time is similar to that for 500nm, 575nm and 800nm excitations. From the ΔN/N values presented, we have only about 1% variation in the carrier density due to pump excitation at the L-symmetry point. The dipole moment estimated for this process is $1.53 \pm 0.03 \times 10^{-31}$ C.m.

Table 3: Excitation wavelength- 650nm, Probe wavelength-800nm

| Excitation power (mJ/cm$^2$) | ΔT Rise time (ps) | ΔT decay time (ps) | ΔR Rise time (ps) | ΔR decay time (ps) |
|---|---|---|---|---|
| 14 | 0.487±0.01 | 1.818±0.02 | 0.286±0.01 | 1.73±0.03 |
| 11 | 0.454±0.01 | 1.471±0.02 | 0.231±0.01 | 1.59±0.02 |
| 7 | 0.519±0.02 | 1.163±0.02 | 0.214±0.01 | 1.28±0.02 |
| 4 | 0.442±0.02 | 1.054±0.02 | 0.214±0.01 | 1.10±0.03 |
| 1 | 0.529±0.05 | 0.967±0.04 | 0.191±0.02 | 1.03±0.04 |

**800nm pump:**

In the last case, both pump and probe wavelengths are set at 800nm. The decay time of differential transmission and reflection for 10 mJ/cm$^2$ pump fluence was found to be $1.61 \pm 0.06$ ps and $1.11 \pm 0.06$ ps, respectively. With decrease in pump power, the differential transmittance and reflectance signals decrease and their decay times are also shorter. The rise and decay times are summarized in Table 4.

The differential transmittance decay time varies linearly with pump fluence. 1.55 eV is below the interband transition energy so the mechanism is, mainly, low energy plasmons leading to intraband excitation of sp- band electrons [49]. The differential reflectance decay times, though linear, has much weaker dependence on the pump fluence. At higher pump powers, measured transmittance and reflectance decay times are significantly different. This is mainly due to smaller (25 nm) skin depth of gold at 800nm [50]. Due to diffusion of hot carriers into the metal, the effective hot carrier distributions are different for transmission and reflection after plasmon excitation. The difference in the decay times of transmittance and reflectance corresponds to the bulk scattering contribution which varied from 120 to 500 fs when pump fluence changed from 1 to 10 mJ/cm$^2$. The dipole moment for this transition is estimated to be $1.73 \pm 0.01 \times 10^{-31}$ C.m.

Table 4: Rise and decay times for excitation wavelength- 800nm, Probe wavelength-800nm

| Excitation power (mJ/cm$^2$) | ΔT Rise time (ps) | ΔT Decay time (ps) | ΔR Rise time (ps) | ΔR Decay time (ps) |
|---|---|---|---|---|
| 10 | 0.28±0.02 | 1.615±0.06 | 0.37±0.05 | 1.11±0.06 |
| 7 | 0.28±0.03 | 1.33±0.080 | 0.5±0.11 | 1.06±0.13 |
| 5 | 0.25±0.05 | 1.323±0.12 | 0.34±0.26 | 0.90±0.19 |
| 2 | 0.2 ± 0.05 | 1.043±0.13 | 0.5±0.1 | 0.92±0.12 |
| 1 | 0.3 ± 0.15 | 0.991±0.1 | 0.55±0.1 | 0.87±0.1 |

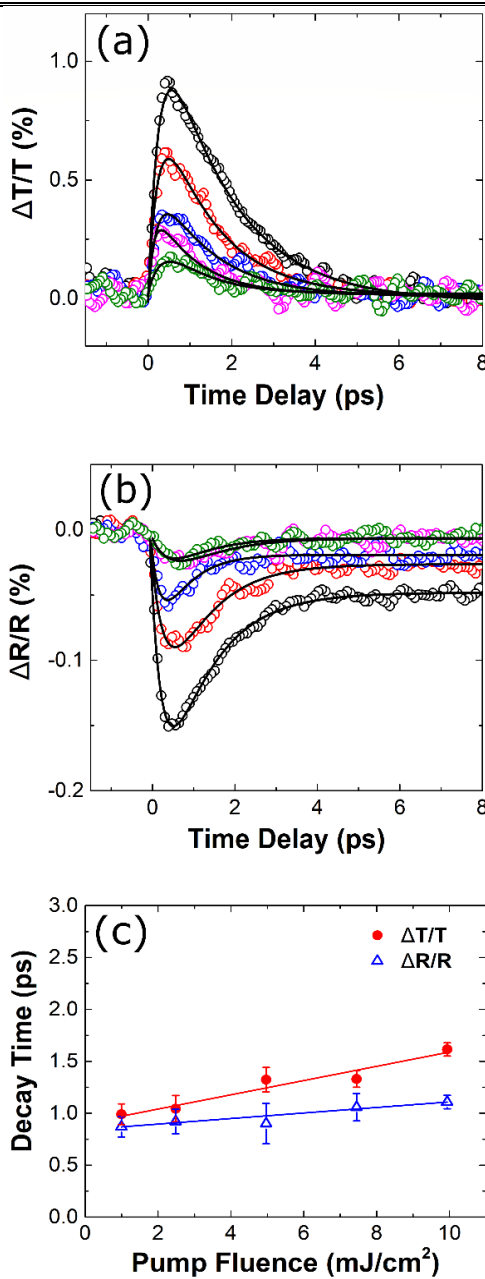

Fig.6 Differential transmission (a) and reflection (b) decay scans at different pump fluences are shown along with fits. Pump wavelength is 800 nm and probe is at 800 nm. (c) shows the pump fluence dependence on the decay time of transmission (Red circles) and reflection (Blue triangles) data. Solid lines show linear fits.

**Decay time at different wavelength excitations:**

Figure 7 shows the interpolated decay times for different excitation wavelengths plotted for various excitation fluences. From the slope of these curves, there is a threshold fluence where the slope changes from positive (at low densities) to negative (at high densities). Even after correcting for the absorption coefficient of gold which increases at the most by a factor of 2 from 500 nm and 800 nm, only the low fluence data will show the decay to be similar for all excitation energies probed here. For the high fluence data (> 10 mJ/cm$^2$), the slope variation would be stronger after correcting for the absorption. That is, the decay times are longer for 800 nm excitation compared to the 500 nm excitation. This can be explained by the screened coulomb potential at higher densities which reduces the scattering cross-section and thus longer decay time. This is further supported by the fact that the change in the decay time as fluence is increased is larger for longer wavelengths as the carrier generation process is more efficient at longer wavelengths due to stronger plasmon mediated intraband carrier generation and higher absorption coefficient of gold. Inset of Fig.7 shows the slope vs fluence along with a straight line fit. From the fit, at the threshold fluence of 9.1 mJ/cm$^2$, the decay time is same at all wavelengths or energies.

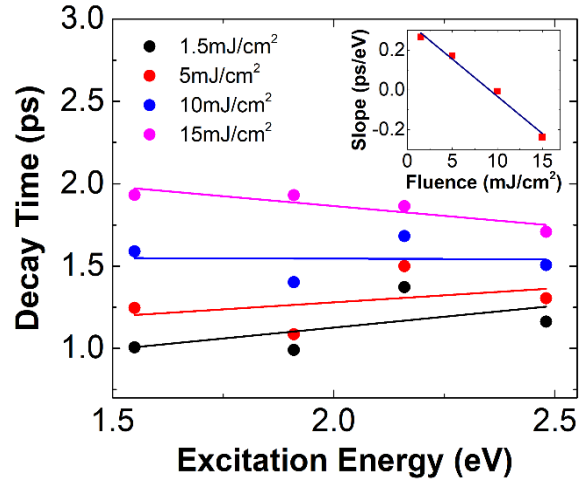

Fig.7 Decay times plotted for different pump excitation energies at different pump fluences are shown along with linear fits. Inset shows the slope vs pump fluence showing a threshold fluence where fluence has no influence on the decay at all energies (slope is zero). Straight line shows linear fit.

**Conclusions:**

In plasmonic structures, absorption is governed by the geometry or structure defined resonances. Non-degenerate and degenerate pump-probe measurements with 45 fs time resolution were carried out on a plasmonic quasicrystal that supports broadband dispersionless plasmons. Three different scenarios were probed to cover how the direct hot carrier excitation in gold, high energy and resonant plasmon excitation affect the plasmon dynamics. We establish the role of direct hot carrier generation as well as the hot carriers generated by hot plasmon decay on the low energy plasmon dynamics.

The characteristics of decay profile depends on excitation wavelength regime as well as pump fluence. The former is strongly correlated to plasmonic structure. We conclude that the decay time show linear dependence on pump fluence for direct hot carrier excitation at 500nm and high energy plasmon excitation at 575nm and resonant plasmon excitation at 800nm. However, it shows a quadratic dependence on pump fluence for plasmonic excitation at 650nm. Quantitatively, at a constant pump fluence, the decay time of transient optical property for low energy plasmons is nearly identical for all scenarios at fluence of 9.1 mJ/cm$^2$. As we get farther from this range towards either end, the decay time varies considerably from one excitation to another. Hence, the density of hot plasmons and their resonance modes affect the plasmon dynamics as both the number of hot carriers generated and the initial hot electron temperature varies from one case to another. The overall variation in the decay time of low energy plasmons obtained by considering all scenarios is in the range of 1- 3 psec.